# Optimization of Quality of Experience through File Duplication in Video Sharing Servers

Emad Abd-Elrahman, Tarek Rekik and Hossam Afifi
Wireless Networks and Multimedia Services Department
Institut Mines-Télécom, Télécom SudParis - CNRS Samovar UMR 5157, France.
9, rue Charles Fourier, 91011 Evry Cedex, France.
{emad.abd_elrahman, tarek.rekik, hossam.afifi}@it-sudparis.eu

## ABSTRACT
Consumers of short videos on Internet can have a bad Quality of Experience (QoE) due to the long distance between the consumers and the servers that hosting the videos. We propose an optimization of the file allocation in telecommunication operators' content sharing servers to improve the QoE through files duplication, thus bringing the files closer to the consumers. This optimization allows the network operator to set the level of QoE and to have control over the users' access cost by setting a number of parameters. Two optimization methods are given and are followed by a comparison of their efficiency. Also, the hosting costs versus the gain of optimization are analytically discussed.

## Categories and Subject Descriptors
C.2.1 [COMPUTER-COMMUNICATION NETWORKS]: Network Architecture and Design.

## General Terms
Algorithms, Performance.

## Keywords
Server Hits; File Allocation; File Duplication; Optimization; Access Cost.

## 1. INTRODUCTION
The exponential growth in number of users access the videos over Internet affects negatively on the quality of accessing. Especially, the consumers of short videos on Internet can perceive a bad quality of streaming due to the distance between the consumer and the server hosting the video. The shared content can be the property of web companies such as Google (YouTube) or telecommunication operators such as Orange (Orange Video Party). It can also be stored in a Content Delivery Network (CDN) owned by an operator (Orange) caching content from YouTube or DailyMotion. The first case is not interesting because the content provider does not have control over the network, while it does in the last two cases, allowing the network operator to set a level of QoE while controlling the network operational costs.

Quality of Experience (QoE) is a subjective measure of a customer's experiences with his operator. It is related to but differs from Quality of Service (QoS), which attempts to objectively measure the service delivered by the service provider. Although QoE is perceived as subjective, it is the only measure that counts for customers of a service. Being able to measure it in a controlled manner helps operators understand what may be wrong with their services to change.

There are several elements in the video preparation and delivery chain, some of them may introduce distortion. This causes the degradation of the content and several elements in this chain can be considered as "QoE relevant" for video services. These are the encoding systems, transport networks, access networks, home networks and end devices. We will focus on the transport networks behavior in this paper.

In this work, we propose two ways of optimization in file duplication either *caching* or *fetching* the video files. Those file duplication functions were tested by feeding some YouTube files that pre-allocated by the optimization algorithm proposed in the previous work [1].

By caching, we mean duplicate a copy of the file in different or another place than its original one. While, in fetching, we mean retrieve the video to another place or zone in order to satisfy instant needs either relevant to many requests or cost issues from the operators point of view.

The file distribution problem has been discussed in many works. Especially the multimedia networks, the work in [2] handled the multimedia file allocation problem in terms of cost affected by network delay. A good study and traffic analysis for inter-domain between providers or through social networks like YouTube and Google access has been conducted in [3]. In that work, there is high indication about the inter-domain traffic that comes from the CDN which implies us to think in optimizing files allocations and files duplications.

Akamai [4] is one of the more famous media delivery and caching solution in CDN. They proposed many solutions for media streaming delivery and enhancing the bandwidth optimization for video sharing systems.

The rest of this paper is organized as follows: Section 2 presents the stat-of-the-art relevant to video caching. Section 3 highlights allocation of files based on the number of hits or requests to any file. In section 4, we propose the optimization by two file duplication mechanisms either caching or fetching. Numerical results and threshold propositions are introduced in Section 5. The hosting issues and gain from this optimization are handled by Section 6. Finally, this work is concluded in Section 7.

## 2. STAT-OF-THE-ART
Caching algorithms were mainly used to solve the problems of performance issues in computing systems. Their main objective was to enhance computing speed. After that, with the new era of multimedia access, the term caching used to ameliorate the accessing way by caching the more hits videos near to the clients. With this aspect, the Content Delivery Network CDN appeared to manage such types of video accessing and enhance the overall performance of service delivery by offering the content close to the consumer.

For the VOD caching, the performance is very important as a direct reflection to bandwidth optimization. In [5], they considered the caching of titles belonging to different video services in IPTV network. Each service was characterized by the number of titles, size of title and distribution of titles by popularity (within service) and average traffic generated by subscribers' requests for this service. The main goal of caching was to reduce network cost by serving maximum (in terms of bandwidth) amount of subscribers' requests from the cache. Moreover, they introduced the concept of "cacheability" that allows measuring the relative benefits of caching titles from different video services. Based on this aspect, they proposed a fast method of partition a cache optimally between objects of different services in terms of video files length. In the implementation phase of this algorithm, different levels of caching were proposed so as to optimize the minimum cost of caching.

The work presented in [6] had considered different locations of doing caching in IPTV networks. They classified the possible locations for caching to three places (STB, aggregated network (DSLAMs) or Service Routers SRs) according to their levels in the end-to-end IPTV delivery. Their caching algorithm takes the decisions based on the users' requests or the number of hits per time period. Caching considers different levels of caching scenarios. But, this work considered only the scenarios where caches are installed in only a single level in the distribution tree. In the first scenario each STB hosts a cache and the network does not have any caching facilities while, in the second and the third scenarios the STBs do not have any cache, but in the former all caching space resides in the DSLAMs, while in the latter the SRs have the only caching facility. The scenario where each node (SR, DSLAM and STB) hosts a cache and where these caches cooperate was not studied in that work. Actually, the last scenario could be complicated in the overall management of the topology. Finally, their caching algorithm was considered based on the relation between Hit Ratio (HR) and flows Update Ratio (UR) in a specific period of time.

The same authors presented a performance study about the caching strategies in on-demand IPTV services [7]. They proposed an intelligent caching algorithm based on the object popularity in context manner. In their proposal, a generic user demand model was introduced to describe volatility of objects. They also tuned the parameters of their model for a video-on-demand service and a catch-up TV service based on publicly available data. To do so, they introduced a caching algorithm that tracks the popularity of the objects (files) based on the observed user requests. Also, the optimal parameters for this caching algorithm in terms of capacity required for certain hit rate were derived by heuristic way to meet the overall demands.

Another VOD caching and placement algorithm was introduced in [8]. They used heuristic ways based on the file history for a certain period of time (like one week) to guide the future history and usage of this file. Also, the decisions were made based on the estimations of requests of specific video to a specific number as a threshold value. The new placements are considered based on the frequency of demands that could update the system rates or estimated requests. Finally, their files distributions depend on the time chart of users' activities during a period of time (for example one week) and its affections on files caching according to their habits. To test this algorithm, they used traces from operational VOD system and they had good results over Least Recently Used (LRU) or Least Frequently Used (LFU) algorithms in terms of link bandwidth for caching emplacements policies. So, this approach is considered as a rapid placement technique for content replications in VOD systems.

An analytical model for studying problems in VOD caching systems was proposed in [9]. They presented hierarchical cache optimization (i.e. the different levels of IPTV network). This model depends on several basic parameters like: traffic volume, cache hit rate as a function of memory size, topology structure like DSLAMs and SRs and the cost parameters. Moreover, the optimal solution was proposed based on some assumptions about the hit rates and network infrastructure costs. For example, the hit rate is a function of memory used in cache and there is a threshold cost per unit of aggregation networks points like DSLAMs. Also, they demonstrated different scenarios for optimal cache configuration to decide at which layer or level of topology.

A different analysis introduced in [10] about data popularity and affections in videos distributions and caching. Through that study, they presented an extensive data-driven analysis on the popularity distribution, popularity evolution, and content duplication of user-generated (UG) video contents. Under this popularity evolution, they proposed three types of caching:

- *Static caching:* at starting point of cache, handling only long-term popularity
- *Dynamic caching:* at starting point of cache, handling the previous popularities and the requests coming after the starting point in the period of trace
- *Hybrid caching:* same as static cache but with adding the most popular videos in a day

By simulation, the hybrid cache improved the cache efficiency by 10% over static one. Finally, this work gave complete study about popularity distribution and its correlations to files allocations or caching.

We will now start by analyzing the optimization of the file allocation introduced in [1] in order to reduce the total access cost, taking into account the number of hits on the files. Then, we will analyze the file duplication algorithms.

## 3. OAPTIMIZATION OF FILES BEST LOCATION

In order to reduce the total access cost, we first move files so as to have every one of them located in the node from where the demand is the highest, i.e. where it has the greatest number of hits. We can follow the steps of the algorithm in Table 1. The main objective from this algorithm is to define the best allocation zone i servers of file f uploaded from any geographical zone i where (i=1:N) zones.

**Table 1. Best location algorithm**

```
Input N            //number of zones
Input M            //number of servers
Input h_ij         //number of hits from zone i to server j for a specific file
Input d_ij         //the assumed cost between zone i and server j
     Total cost C_i = ∑ h_ij * d_ij        // where i from 1 to N and j from 1 to M
     If C_i < C_{i+1},…,C_N
            Then allocate this file in zone i
            Else move this to min (C_{i+1},…,C_N) value location
     End If
Return the best location for this file     //best j
End
```

We applied that algorithm on six YouTube files [1] chosen from different zones as shown in (Table 2). The numbers are rounded for better clarity. Those files were preselected as the most hit files from different zones analyzed briefly in [1].

**Table 2. Examples of files chosen from YouTube and related hits**

| | Hits | | | | | |
|---|---|---|---|---|---|---|
| | File 1 | File 2 | File 3 | File 4 | File 5 | File 6 |
| Zone 1 | 135000 | 16000 | 55000 | 8700 | 3800 | 42000 |
| Zone 2 | 40000 | 5000 | 175000 | 2700 | 1000 | 15000 |
| Zone 3 | 132000 | 51000 | 55000 | 8800 | 3900 | 43000 |
| Zone 4 | 100000 | 12000 | 40000 | 6400 | 2800 | 138000 |
| Zone 5 | 12000 | 1000 | 5000 | 800 | 12400 | 4000 |
| Zone 6 | 17000 | 1000 | 5000 | 900 | 400 | 5000 |

We reach to the following results of files' distribution or best allocation:

The best location of File 1 is the servers in zone 1 where the algorithm gave the minimum cost; File 2 moves from zone 2 to zone 3; File 3 moves from zone 3 to zone 2; File 4 moves from zone 4 to zone 3: File 5 stays in zone 5; File 6 moves to zone 4.

The new files distribution is shown in Figure 1.B and will be the distribution that we'll depend on for the next optimization (Duplication).

Since geography plays an important role in delivering video to customers, we propose the following network representation shown in Figure 1. (A). We divide the network into six zones each zone represented by a node and we select a file uploaded from each zone to be studied by our algorithms.

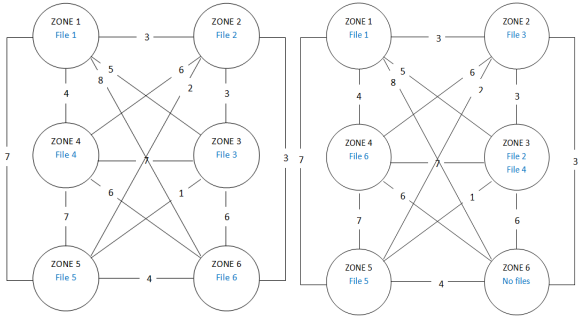

Figure 1. (A) Network Representation    (B) File distribution after the application of the best location algorithm

The nodes in this figure represent the servers of a geographical zone; it can be a city for a national operator. A node can also be assimilated to the edge router connecting the customers who live in that geographical zone to the operator's core network or (CDN). The arcs connecting the nodes are physical links, and the numbers on them represent an access cost, i.e. the cost of delivering a video from zone A to a consumer from zone B. That access cost (is assumed to be symmetric) can be a combination of many parameters such as the path's length, the hop count (from edge router to edge router), the cost of using the link (optical fiber, copper, regenerators and repeaters, link rental cost) and the available bandwidth. The files of a given zone are hosted by the servers of that zone.

## 4. OPTIMIZATION THROUGH FILE DUPLICATIONS

To grant a given QoE, we choose to give access to popular file (located in zone **j** servers and having a number of hits higher than a given value **Y**) for a given consumer (from zone **i**) only if this consumer can access that file with a cost $a_{ij}$ lower than a given value **A** ($a_{ij}$ =< **A**). If it is the case, we check the demand on that same file from the zone **i** of the customer (which is $y_{ij}^f$). If the demand is higher than a threshold **Y**, we duplicate file **f**. Then, any consumer can access any content with good quality and minimum cost. The steps of file duplication algorithm are shown in Table 3.

Thus, we make a compromise between access cost and storage cost. Actually, the higher the value of **Y**, the higher the access cost and the lower the need for storage capacity. Also, the higher the value of **A**, the higher the access cost and the lower the need for storage capacity. The values of **A** and **Y** are thresholds being adjusted by the operators (Content Delivery).

We are going to experiment two different duplication functions:

- *duplicate1*(f,i,j) duplicates file *f* (located in zone j) in zone i which means *(caching)* Table 4.

- *dupliacate2*(f,i,j) duplicates file *f* in the zone **k** that has the second highest number of hits on that file (located in zone j) and that grants an access cost lower than **A** at the same time, zone j being the zone with the first highest number of hits (the best location). This means *(fetching)* Table 5.

We have to make sure that the duplicated files are not taken into consideration while running the algorithm in Table 3, i.e. in line "For f from 1 to $p_j$" *f* can't be a duplicated file.

We apply the duplication algorithm (Table 3) to the 6 files, starting from the file distribution on Figure 1.B where every file is located in its best location.

We are going to look at different values of Y (0, 5000, 10000 and 20000) and compare the efficiency of the 2 duplicating methods for each value of Y. Then, we will also see the effect of changing Y value on the total gain associated with the optimization. For all cases of Y, we suppose A=5 unit cost.

**Table 3. Duplication algorithm**

```
Input N          // number of zones
Input A          // maximum access cost, set by the network operator
Input Y          // minimum number of hits allowing file duplication,
                 //set by the network operator

Input a_ij       // access cost to a file located in zone j from a customer located in zone i
Input p_j        // number of files in zone j
Input y_ij^f     // number of hits on file f (located in zone j) from zone i users

For i from 1 to N                         //user's zone
    For j from 1 to N                     //file's zone
        If a_ij > A                       // we suppose that a_ii=0
            Then For f from 1 to p_j
                If y_ij^f > Y
                    Then duplicate(f,i,j) // duplicates file f
                                          //in the best location
                End If
        End If
End
```

**Table 4. Function 'duplicate1'**

```
Function duplicate1(f,i,j)
    Duplicate file f in zone i
End
```

**Table 5. Function 'duplicate2'**

```
Function duplicate2(f,i,j)

Input k=j                              // zone k is the best location to duplicate file f
While a_ik > A
    k:=retrieve(f,i)
Duplicate file f in zone k             // we suppose that there's at least one zone k
                                       // that ensures a_ik =< A
End
```

## 5. METHODS AND NUMERICAL RESULTS

This section focuses on performance investigation of the proposed methods for files duplication by applying these two functions on different files from YouTube. Moreover, we will test different use cases by changing the threshold value Y that supposes to be adjusted by the operator as follows:

### 5.1 First Case Y=0

We get the following file distribution with *'duplicate1'*. Assume **'o'** indicates the file's best location, while **'x'** indicates the file has been duplicated as in Table 6.

Let's now compute the gain achieved through the first duplication method, i.e. the difference between the access costs before and after the application of the Duplication algorithm, $a_{ij}^f$ being the access cost to File f located in zone j from a consumer located in zone i.

**Table 6. File distribution with Y=0 & caching**

|        | Zone 1 | Zone 2 | Zone 3 | Zone 4 | Zone 5 | Zone 6 |
|--------|--------|--------|--------|--------|--------|--------|
| File 1 | o      |        |        |        | x      | x      |
| File 2 |        |        | o      | x      |        | x      |
| File 3 |        | o      |        | x      |        |        |
| File 4 |        |        | o      | x      |        | x      |
| File 5 | x      |        |        | x      | o      |        |
| File 6 |        | x      | x      | o      | x      | x      |

*Example: Gain from the duplication of File 5 in zone 1*

- File 5 is no longer delivered to zone 1 consumers from zone 5 but from zone 1.

  Gain from zone 1 = $(a_{15} - a_{11}) * y_{15} f_5 = (7-0)*3800 = 26600$

- Consumers from zone 2 still access File 5 that is located in zone 5 because it is the zone with the cheapest access cost ($a_{25}=2$) compared with the other zones that host File 5 (zones 1 and 4 for which the access costs from zone 2 are respectively 3 and 6). Gain from zone 2 = 0

- Consumers from zone 3 still access File 5 that is located in zone 5 because it is the zone with the cheapest access cost ($a_{35}=1$) compared with the other zones that host File 5 (zones 1 and 4 for which the access costs from zone 3 are respectively 5 and 7). Gain from zone 3 = 0

- Consumers from zone 4 access File 5 that is located in zone 4 because it is the zone with the cheapest access cost ($a_{44}=0$). Gain from zone 4 = 0

- Consumers from zone 5 access File 5 that is located in zone 5 because it is the zone with the cheapest access cost ($a_{55}=0$). Gain from zone 5 = 0

- Consumers from zone 6 still access File 5 that is located in zone 5 because it is the zone with the cheapest access cost ($a_{65}=4$) compared with the other zones that host File 5 (zones 1 and 4 for which the access costs from zone 6 are respectively 8 and 6). Gain from zone 6 = 0

We repeat the same operation for the other duplicated files and we get the following file distribution with the fetching method *(duplicat2)*. This distribution is shown in Table 7 with **'o'** indicates the file's best location, while **'x'** indicates the file has been duplicated.

**Table 7. File distribution with Y=0 & fetching**

|        | Zone 1 | Zone 2 | Zone 3 | Zone 4 | Zone 5 | Zone 6 |
|--------|--------|--------|--------|--------|--------|--------|
| File 1 | o      | x      | x      |        |        |        |
| File 2 | x      | x      | o      |        |        |        |
| File 3 | x      | o      |        |        |        |        |
| File 4 | x      | x      | o      |        |        |        |
| File 5 | x      |        | x      |        | o      |        |
| File 6 |        | x      | x      | o      |        |        |

We notice that with caching, the majority of the duplicated files are distributed among zones 4, 5 and 6, while with fetching zones 1, 2 and 3 contain all the duplicated files.

After computing the new access costs, we compare the gain achieved through the duplicating methods in the figures below (Figure 3).

If we look at File 1, we notice that the sum of the access costs from all the zones after caching (the sum of the red columns) is higher than that after fetching (the sum of the green columns). Therefore, there's a greater gain with fetching than with caching to access File 1 from all over the network.

After computing the new access costs, we compare the gain achieved through the duplicating methods in the figures below (see Figure 2).

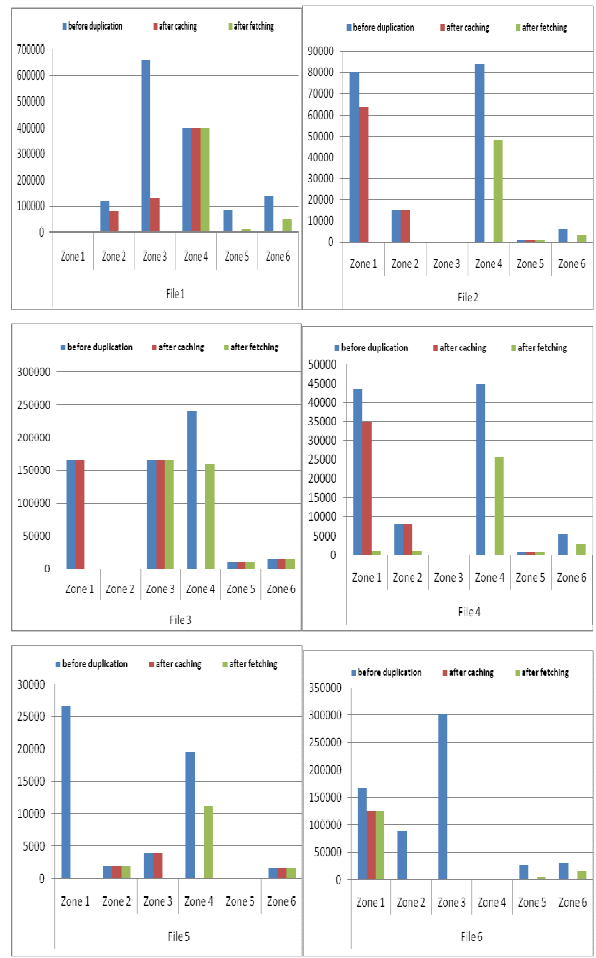

Figure 2. Access cost before and after *duplication1* and *duplication2* for Y=0

For the delivery of File 1 to zone 4, no gain was achieved. That means zone 4 customers still access File 1 from the same zone (zone 1 here). In other cases, there may be a duplication of that file on another zone that has the same access cost to zone 4 as zone 1 (the best location of File 1).

We notice that fetching is more efficient than caching for Files 1, 2, 3 and 4 and zones 1, 2 and 3, while caching is better for Files 5 and 6 and zones 4, 5 and 6.

### 5.2 Second Case Y=5000

For this threshold, we notice that with caching algorithm, the majority of the duplicated files are distributed among zones 4, 5 and 6, while with fetching algorithm, zones 1, 2 and 3 contain all the duplicated files, like for Y=0.

In Figure 3 below, we only show the files that were duplicated by any of the 2 duplication methods.

We notice that fetching is more efficient than caching for Files 1, 2, 3 and 4 and zones 1, 2 and 3 (except for File 6), while caching is better for File 6 and for zones 4, 5, 6, even if the two methods are of equal efficiency on certain zones.

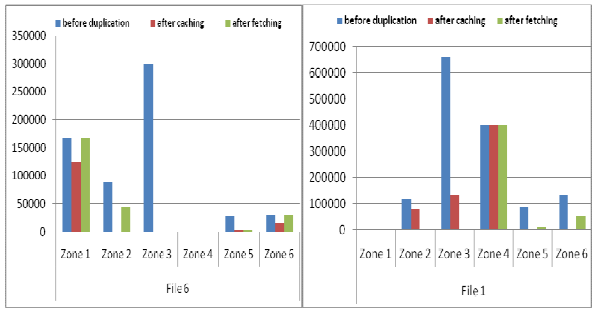

Figure 3. Access cost before and after caching and fetching for Y=5000

We also notice that File 5 was not duplicated because there is no enough demand for it. The only zone from which the demand exceeds the demand threshold Y (5000) is zone 5 (which is the best location of File 5) with 12400 hits (see Table 2).

**5.3 Third Case Y=10000**
After applying the two methods, we get the following files distribution either for caching or for fetching:

Like for Y=0 and Y=5000, we notice that with caching, the majority of the duplicated files are distributed among zones 4, 5 and 6, while with fetching zones 1, 2 and 3 contain all the duplicated files.

This is understandable for fetching as the highest demands are those of Files 1, 2 and 3, and these files are located in zones 1, 3 and 2 respectively. It is also understandable for caching as the access cost from zones 4, 5 and 6 to Files 1, 2 and 3 is high compared to access costs between zones 1, 2 and 3:

- Zone 4 customers access File 3 with an access cost of 6, and File 2 with 7;
- Zone 5 customers access File 1 with an access cost of 7;
- Zone 6 customers access File 1 with an access cost of 8 and File 2 with 6.

In Figure 4, we only show some files that were duplicated by any of the 2 duplication methods which are files 3 and 6.

We notice that fetching is more efficient than caching for Files 1, 2 and 3 and zones 1, 2 and 3 (except for File 6), while caching is better for File 6 and for zones 4, 5, 6, even if the two methods are of equal efficiency on certain zones.

We also notice neither File 4 nor File 5 was duplicated because there's not enough demand for them. The highest demand for File 4 is 8800 (as shown in Table 2) and doesn't exceed Y (10000), and the only zone from which the demand exceeds Y for File 5 is zone 5 (which is the best location of File 5) with 12400 hits (see Table 2).

**5.4 Forth case Y=20000**
We get the following file distribution for both file duplication methods as shown below.

In the Figure 5, we only show sample of files that were duplicated by any of the two duplication methods which are files 3 and 6.

We notice that fetching is more efficient than caching for File 3 and zones 1 and 2, while it has the same efficiency as caching on File 6 and zones 3, 5 and 6. Caching is more efficient than fetching only on zone 4.

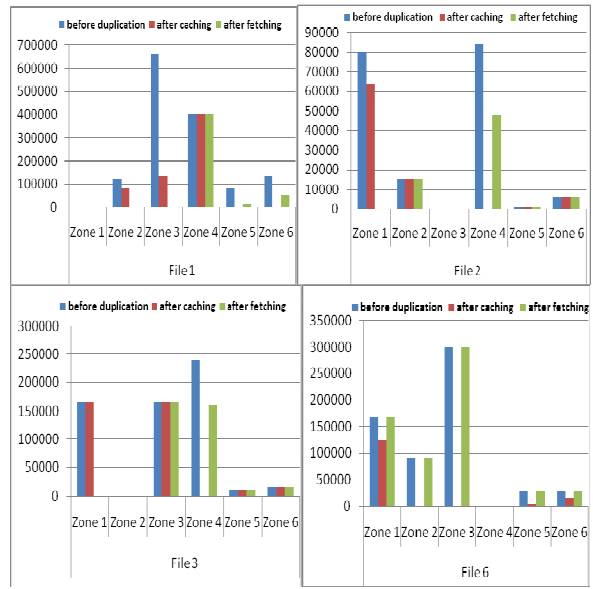

Figure 4. Access cost before and after caching and fetching for Y=10000 to files 1,2,3 and 6

Here, despite the fact that there is a demand higher than Y (20000) on File 1 from zones other than its best location (zone 1), such as zone 3 (132000 hits), zone 4 (100000 hits) or zone 2 (40000 hits), File 1 was not duplicated. This is due to the access cost threshold A for zones 2, 3 and 4, and due to the demand threshold Y for zones 5 and 6 as there is not enough demand on File 1 from these 2 zones (12000 and 17000 respectively).

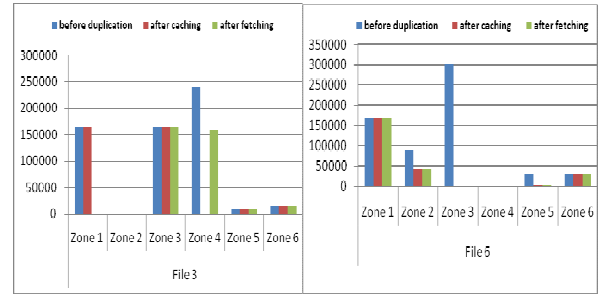

Figure 5. Access cost before and after caching and fetching for Y=20000 to files 3 and 6

## 6. HOSTING COST AND NET GAIN
In this part, we take into account the file duplication cost. This cost is mainly due to file hosting. We assume that:

- A hit cost is 0.01 m.u. (money unit). Then, we multiply the access costs by 100 to have all the costs expressed in m.u. (just to scale the values) ;
- 1 TB (unit size in Bytes) hosting cost is 20 m.u. ;
- The files are sets of files, each set is 100 TB; in fact, the hosting servers contain a great number of videos, and the network operator may duplicate sets of files instead of running the Duplication algorithm for every single file. These sets may encompass files that have almost the same viewers (all the episodes of a TV show for example).

The hosting cost = number of duplicated files * file size in TB * 1 TB hosting cost.

**Table 8. Number of duplicated files with both duplication methods**

| Y | 0 | 5000 | 10000 | 20000 |
|---|---|------|-------|-------|
| Number of duplicated files through caching | 13 | 7 | 6 | 2 |
| Number of duplicated files through fetching | 11 | 5 | 4 | 2 |

We notice that the number of duplicated files decreases if Y increases. This is due to the fact that only the most popular files are duplicated with a high value of Y as shown in Table 8.

Now if we compare the gain in access cost and the hosting cost (a shown Figure 6), we notice that there is a financial loss for Y=0 with both duplication methods. This is understandable as duplicating files that are not popular enough require important hosting resources and benefit to a small number of customers. Hence the need to compute the Net gain, which is the difference between the gain in access cost and the hosting cost, in order to properly assess the efficiency of both duplication methods.

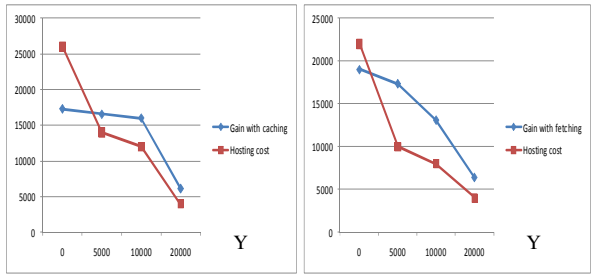

Figure 6. Gain vs. hosting cost

With caching, there is a gain starting from Y≈4000 and the biggest Net gain is achieved for Y=10000 according to Figure 8. While with fetching, there is a gain starting from Y≈2000 and the biggest net gain is achieved for Y=5000 according to the same Figure 7.

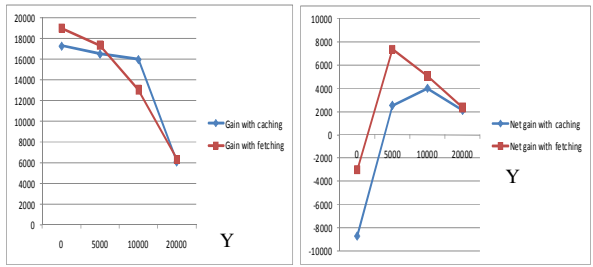

Figure 7. Gain and Net gain comparison

We notice that fetching is more efficient in terms of Net gain, despite the fact that caching is better for Y between 6000 and 19000 in terms of Access gain. This is due to the fact that the duplicated files are smartly allocated with fetching, allowing more customers to access them with a minimum cost, and because there's a lower need for duplicating files than with caching.

We also notice that beyond a certain value of Y, the Net gain decreases, and can even become negative beyond a certain limit (> 20000 in our example). Moreover, as Y increases beyond 10000, the advantage of fetching over caching diminishes.

## 7. CONCLUSION

This paper provided a study for video files distribution and allocation. We started by highlighting some techniques used in video files caching and distributions. Then, we proposed two mechanisms of files duplications based on files hits and some access cost assumptions set by the operators.

Despite the fact that fetching is better in terms of Net gain, both duplication methods can be more or less efficient, as we have seen that caching is more efficient for the delivery of File 6 for Y=0, 5000 or 1000, and has equal efficiency with fetching for Y=20000. We have also seen that, setting the right demand threshold Y is a key step to achieve cost optimization, along with the access cost threshold A (A=5 in this paper) that we didn't discuss.

We can also combine duplication methods or use new ones for better results. We can, for example, duplicate a file in the zone that has the second and the third highest number of hits on that file and that grants an access cost lower than A at the same time.

However, the results found in the treated example may vary depending on the network configuration (the access costs, the number of files and their distribution) and a duplication method that proves to be efficient for a given network may not work for another one.

Finally, if we compare our proposed techniques with the previous stated works we can find that our algorithms are heuristic ones. This means that, the correlation made between user requests and the network conditions (set by the operator) play an important role in caching decisions.